\documentclass[lettersize,journal]{IEEEtran}
\usepackage{amsmath,amsfonts}
\usepackage{algorithmic}
\usepackage{array}
\usepackage[caption=false,font=normalsize,labelfont=sf,textfont=sf]{subfig}
\usepackage{textcomp}
\usepackage{stfloats}
\usepackage{url}
\usepackage{verbatim}
\usepackage{graphicx}
\def\BibTeX{{\rm B\kern-.05em{\sc i\kern-.025em b}\kern-.08em
    T\kern-.1667em\lower.7ex\hbox{E}\kern-.125emX}}
\usepackage{balance}

\usepackage{multicol}
\usepackage{multirow}
\usepackage{lipsum}
\usepackage[capitalise]{cleveref} 
\usepackage[
backend=biber,
style=ieee,
]{biblatex}
\usepackage[dvipsnames]{xcolor}
\usepackage[flushleft]{threeparttable}
\newcolumntype{L}{>{\centering\arraybackslash}m{1.5cm}}
\newcolumntype{l}{>{\centering\arraybackslash}m{1.2cm}}

\begin{document}
\title{Supervised Contrastive ResNet \\and Transfer Learning for \\the In-vehicle Intrusion Detection System}
\author{Thien-Nu Hoang,~\IEEEmembership{Student Member,~IEEE,}
        and Daehee Kim,~\IEEEmembership{Member,~IEEE}
\thanks{Authors are with the Department
of Future Convergence Technology, Soonchunhyang University, Asan, Republic of Korea}
\thanks{E-mail: htnu@sch.ac.kr, daeheekim@sch.ac.kr.}
\thanks{Manuscript created October, 2020.}}

\markboth{Journal of \LaTeX\ Class Files,~Vol.~18, No.~9, September~2020}%
{How to Use the IEEEtran \LaTeX \ Templates}

\maketitle

\begin{abstract}
High-end vehicles have been furnished with a number of electronic control units (ECUs), which provide upgrading functions to enhance the driving experience. The controller area network (CAN) is a well-known protocol that connects these ECUs because of its modesty and efficiency. However, the CAN bus is vulnerable to various types of attacks. Although the intrusion detection system (IDS) is proposed to address the security problem of the CAN bus, most previous studies only provide alerts when attacks occur without knowing the specific type of attack. Moreover, an IDS is designed for a specific car model due to diverse car manufacturers. In this study, we proposed a novel deep learning model called supervised contrastive (SupCon) ResNet, which can handle multiple attack identification on the CAN bus. Furthermore, the model can be used to improve the performance of a limited-size dataset using a transfer learning technique. The capability of the proposed model is evaluated on two real car datasets. When tested with the car hacking dataset, the experiment results show that the SupCon ResNet model improves the overall false-negative rates of four types of attack by four times on average, compared to other models. In addition, the model achieves the highest F1 score at 0.9994 on the survival dataset by utilizing transfer learning. Finally, the model can adapt to hardware constraints in terms of memory size and running time.
\end{abstract}


\section{Introduction}\label{sec:intro}


%
%
%
%

The intelligent vehicle industry has gained considerable attention and interest from companies, researchers, and consumers. Many electronic control units (ECUs) are installed inside a smart vehicle to assist with more advanced functions. These ECUs are interconnected through an in-vehicle network (IVN) in which the controller area network (CAN) protocol is widely used, although other technologies, such as the local interconnected network (LIN), FlexRay, and Ethernet, are also available. The CAN protocol lacks encryption and authentication mechanisms despite its fast speed and simplicity. Hence, there is a trade-off between efficiency and security. Many studies demonstrate that a CAN bus network can be attacked in various ways \cite{Koscher2010ExperimentalAutomobile, Hoppe2011SecurityCountermeasures, Jo2021ACountermeasures}. Thus, different mechanisms have been introduced to fill the gap in the security of the CAN bus. One of them is developing a system monitoring and detecting attacks in the CAN bus network, which is called an intrusion detection system (IDS). With the rapid development of the machine learning field, various studies applied simple machine learning models, including K-means \cite{DAngelo2020AVehicles}, K nearest neighbors (KNN) \cite{Derhab2021Histogram-BasedNetworks}, and one-class support vector machine (OSVM) \cite{Avatefipour2019AnLearning}, or used complex deep learning models, such as convolutional neural network (CNN) \cite{Song2020In-vehicleNetwork, Desta2022Rec-CNN:Plots} and long-short term memory (LSTM) \cite{Taylor2016AnomalyNetworks, Ashraf2020NovelSystems, Nam2021IntrusionNetworks} to build an efficient IDS. Although the previous studies have achieved noticeable results with high detection accuracy and low error rate, there are still two main issues. First, most of them solve the problem of binary classification, which consists of two classes: normal and abnormal. In this study, we aim to solve the multiclass classification that distinguishes normal traffic and different types of attacks, which will then notify the user of the specific type of attack. Thus, we propose the concept of contrastive learning \cite{ChopraLearningVerification}, which learns from dissimilarity and similarity between training samples because multiclass classification is more challenging than binary classification. In addition, contrastive learning-based models can solve the challenge of class imbalance in the CAN bus dataset. Second, previous machine learning-based IDS learns the behavior of CAN message transmissions of a specific vehicle model. Further, CAN ID meanings are distinct for different vehicle models although the CAN messages follow the same structure. Therefore, it cannot be applied to other vehicle models that require the collection of a new large dataset and training of a new model if we want to develop a new efficient IDS for a newly launched vehicle model. Hence, we apply the transfer learning technique to address the problem in this study.
The main contributions of this study can be summarized as follows:
\begin{itemize}
\item We introduce a novel model called supervised contrastive (SupCon) ResNet, which combines supervised contrastive learning with the traditional ResNet and minimizes the supervised contrastive loss, rather than the traditional cross-entropy loss. To the best of our knowledge, our study is the first to apply contrastive learning to the in-vehicle IDS. 
\item We build a domain-adaptation transfer learning system, which facilitates the generalization of a deep learning model across different CAN bus data from different car models. We used the pre-trained SupCon ResNet model to transfer learned knowledge from a rich source dataset to another target dataset. The experiment results show that the proposed SupCon-based transfer learning system lowers the false-negative rate of the training from scratch approach and CE-based transfer learning by 0.4\% and 0.2\%, respectively.
\item We prove the efficiency of our proposed system through comprehensive experiments and analyses on two popular real car datasets: the car hacking \cite{Song2020In-vehicleNetwork} and survival \cite{Han2018AnomalyAnalysis} datasets. The results show that our proposed model is superior to other baselines in terms of detection performance (obtained the highest F1 score of 0.9998), memory size, and running time.
\end{itemize}
After providing an overview of the problem and stating the contributions of the study in \cref{sec:intro}, the relevant background knowledge and security issues in the CAN bus protocol are introduced in \cref{sec:background}. In \cref{sec:related_works}, we summarize various studies related to the in-vehicle IDS. Our proposed system is described in detail in \cref{sec:method}. Next, the experimental setup and results are presented in \cref{sec:results}, where we showed the efficiency of the proposed method empirically. Finally, \cref{sec:conclusion} presents the conclusions and directions for future works. 

\section{Can Bus Background \label{sec:background}}

\begin{figure*}[t]
\centering
\includegraphics[width=7in]{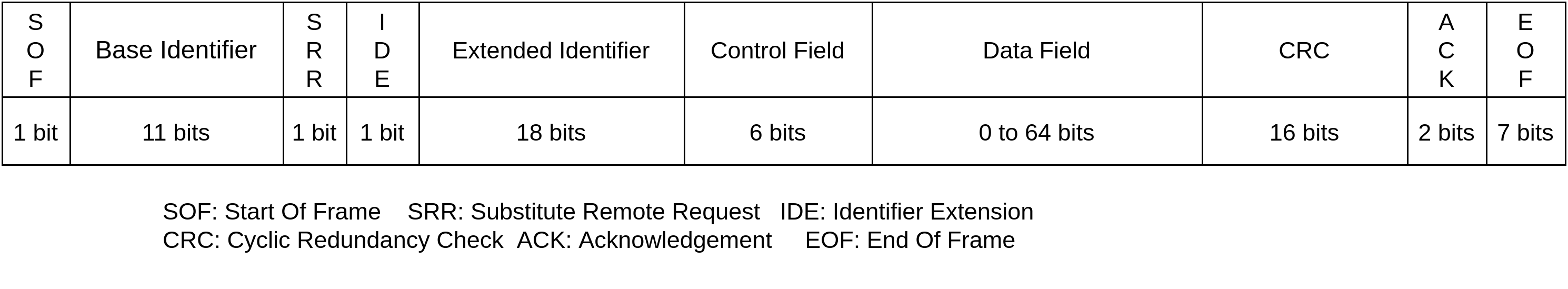}   
\caption{CAN2.0B data frame format.}
\label{fig:can_data_frame}
\end{figure*}
%

\subsection{CAN Bus System}
The controller area network (CAN) is one of the protocols used in the in-vehicle network comprising many connected ECUs. The CAN bus, which is invented by Robert Bosh GmbH in the early 1980s, has become common in automotive systems because of its advantages, including high bit rates (up to 1 Mbit/s), cost-effectiveness, and system flexibility \cite{1991BOSCH2.0}. The concept of CAN protocol comprises message prioritization and multi-master. Every CAN message contains an ID that indicates the content of the message. Based on this, each ECU obtains its relevant messages from a message filterer. There is no source and destination information in CAN messages. In addition, the ID implies the priority of a message: the lower ID, the higher priority. When the bus is idle, any unit can start to transmit a message. The message whose higher priority wins the bus and takes the right of sending the message. During transmission, any number of nodes can receive and simultaneously act upon the same message according to the concept of message filtering.

A CAN message is called a frame belonging to four types, such as data, remote, error, and overload frames. While the data frame contains the main information in ECU communication, the remote frame is used when a unit requests peculiar information from other ECUs. Meanwhile, the error frame supports error detection and fault confinement. An overload frame will be sent whenever a node requires a delay of the next frame. Every CAN frame is made up of a sequence of dominant bits (0) and recessive bits (1). The data and remote frames have the same structure, as shown in \cref{fig:can_data_frame}. The data and remote frames start with a bit called the start of frame (SOF), which is always a dominant bit (0). According to the length of the ID, the CAN data are classified into CAN 2.0A and CAN 2.0B. CAN 2.0A is a shortened version of CAN 2.0B, without the extended identifier. Hence, the CAN 2.0B format structure is described in \cref{fig:can_data_frame}. The 11-bit length base identifier is followed by a 1-bit substitute remote request (SRR) and 1-bit identifier extension (IDE). The successive 18 and 6 bits represent the extended identifier and control field, respectively. The data field containing 64 bits is the content of the message. The following 16 bits for cyclic redundancy check (CRC) and 2 bits for acknowledgment are utilized in error detection and correction. The end of the frame (EOF) is recognized by seven consecutive bits. For more information regarding the other frame types, we refer to the CAN specification reported by Bosh \cite{1991BOSCH2.0}.
\subsection{Security in CAN Bus Protocol}
Despite the advantages mentioned above, the most severe problem in CAN protocol is message transmission without any authentication and encryption. Particularly, any node connected to the bus can obtain the message during message transmission. Consequently, an attacker who compromises a node on the bus can easily sniff the information on the bus. There are two levels of an adversary: weak and strong \cite{Cho2016FingerprintingDetection}.  A weak attacker can prevent the ECU from transmitting messages or keep the ECU in a listen-only mode. Meanwhile, a strong attacker can completely control an ECU and have access to the memory data. Consequently, malicious messages can be injected to launch the attack in addition to the abilities of the weak attacker. These two levels of attacks are proved in several studies \cite{Koscher2010ExperimentalAutomobile, Hoppe2011SecurityCountermeasures, Jo2021ACountermeasures}. In this study, our proposed IDS handles the strong attacker including three different types of injected messages: 
\begin{itemize}
    \item DoS: The attacker floods the bus by injecting messages containing the highest priority ID 0x0000 and arbitrary data fields. The legitimate messages are prevented from being transmitted, resulting in unusual effects, such as flashing dash indicators, intermittent accelerator/steering control, and even full vehicle shutdown, since the message with the ID 0x0000 always wins the bus \cite{Cho2016FingerprintingDetection}. 
    \item Fuzzy: The attacker injects messages with an arbitrary ID and data fields at a high frequency. Aside from being randomly generated, the injected IDs can be chosen from those appearing in the normal traffic, which is supposed to be difficultly detected. The effect of this attack is similar to the DoS attack. 
    \item Target ID: The attacker injects messages with a specific ID and manipulated data fields as his intent. To achieve this, the attack requires the attacker to have knowledge of the meaning of the specific ID by reverse engineering techniques.
\end{itemize}

\section{Related Works \label{sec:related_works}}
\enlargethispage{\baselineskip}
\subsection{Deep Learning-based In-vehicle IDS}
Two main approaches to counter in-vehicle cyber-attacks are message authentication and intrusion detection \cite{Cho2016FingerprintingDetection}. Message authentication consumes more resources and time, although it provides a higher security level. Consequently, these approaches decrease the CAN bus performance. Therefore, intrusion detection is preferred. Different types of IDS have been proposed to monitor and analyze transferred messages on the bus, detect vicious behaviors, and make an alert if detected.

With the rapid development of deep learning models, many deep learning-based in-vehicle IDS have been introduced. In 2016, a previous study in \cite{Kang2016ASecurity} was one of the first research that applied deep learning for CAN bus intrusion detection. The authors developed a deep neural network where each neuron takes each bit of data payload in the CAN message as the input. The network was trained in a supervised manner with two classes: normal and abnormal. However, the proposed model was investigated using a self-synthesized dataset with simple attack models. Although the result was not impressive, this study is the first to apply advanced deep learning models to the CAN bus network. Several studies applied recurrent neural networks (RNN) to identify injected messages since CAN messages possess sequential patterns. In \cite{Taylor2016AnomalyNetworks}, the authors used the long short term memory (LSTM), which is the improvement of RNN, to predict the next bit in the data payload of normal CAN message sequences. The invasion is detected by checking the difference between the predicted and receiving values on the bus. Meanwhile, the LSTM-based autoencoder, which was adopted by \cite{Ashraf2020NovelSystems}, was trained with special statistic features of the CAN message sequences. These models were designed for a message sequence of a specific CAN ID. This implies that we need to train many models separately, corresponding to the number of IDs in a CAN network of a vehicle. To reduce the number of models, the authors in \cite{Nam2021IntrusionNetworks} employed the bi-directional generative pretrained transformer (GPT) network to predict the next CAN ID in the CAN message sequence, which was used for abnormal detection to identify attacks.      

Aside from RNN, the convolutional neural network (CNN) was exploited in many CAN bus IDS studies. CNN is commonly used for various computer vision tasks, such as object detection, image colorization, and image segmentation. Hence, a CNN-based IDS requires an image representation of CAN messages. In \cite{Seo2018GIDS:Network}, multiple one hot vectors representing consecutive CAN IDs were stacked to form a CAN image, which is fed into the CNN generative adversarial networks (GAN). The proposed method can perform binary classification (normal/abnormal) with an overall accuracy of 97\% and detect unknown attacks.  Similarly, a previous study \cite{Song2020In-vehicleNetwork} manipulated CAN ID sequences in binary form, which is used to train a simplified Inception Resnet. The proposed model achieves a low false-negative rate in the case of DoS and spoofing attacks, but the result of the fuzzy attack is not good. Meanwhile, the authors in \cite{Desta2022Rec-CNN:Plots} proposed a concept of recurrence plots, which is the matrix of subtractions value of multiple CAN IDs within a specific window size. In addition, they suggested using a lightweight CNN model with the proposed recurrence plots to foster the speed of IDS. However, the accuracy of multiclass classification is not high.

\subsection{Transfer Learning in the In-vehicle IDS}

All previous models can only be applied to a specific car model. This indicates retraining a new model on a new dataset is required if we build an IDS for a newly released car. In reality, it is time-consuming and requires lots of effort to collect data for every newly released car model. Therefore, transfer learning is proposed to address this problem. As far as we know, the study \cite{Kang2021ASystem} is the first study that employed transfer learning for CAN bus data and presented an LSTM-based model to solve the binary classification. The authors tested their proposed scheme on the survival dataset \cite{Han2018AnomalyAnalysis}, which is a small size dataset. The data of Kia Soul and Chevrolet Spark were used for testing the transfer learning, while the data of Hyundai Sonata car was employed for training the proposed model. However, transfer learning should be used when there is a large source dataset, and the source and target tasks should have the same domain. Therefore, the proposed scheme did not achieve good results.

\subsection{Contrastive Learning in IDS}
To the best of our knowledge, there are no studies applying contrastive learning in the in-vehicle IDS design. However, some recent works have proposed the use of contrastive learning to the network IDS  \cite{Andresini2021Autoencoder-basedDetection, Liu2022ADetection,  Lopez-Martin2022SupervisedDetection}. For example, \cite{Andresini2021Autoencoder-basedDetection} combined the autoencoder and triplet loss to demonstrate an IDS on various network datasets, such as KDDCUP99\footnote{\url{ http://kdd.ics.uci.edu/databases/kddcup99/kddcup99.html}}, AAGM17\footnote{\url{https://www.unb.ca/cic/datasets/android-adware.html}}, and CICIDS17\footnote{\url{https://www.unb.ca/cic/datasets/ids-2017.html}}. Meanwhile, autoencoder-based contrastive learning was introduced as a part of a multi-task model in \cite{Liu2022ADetection}. In addition, \cite{Lopez-Martin2022SupervisedDetection} suggested a novel concept of contrastive learning, in which they projected the labels and features into the same representation space. In the classification phase, the predicted label is a class having the closest distance to the input features in that representation space. All studies mentioned suggested that contrastive learning is suitable to solve the problem of class imbalance, which occurs in IDS research.

\section{Methodology} \label{sec:method}

\begin{figure*}[t]
\centering
\includegraphics[width=5in]{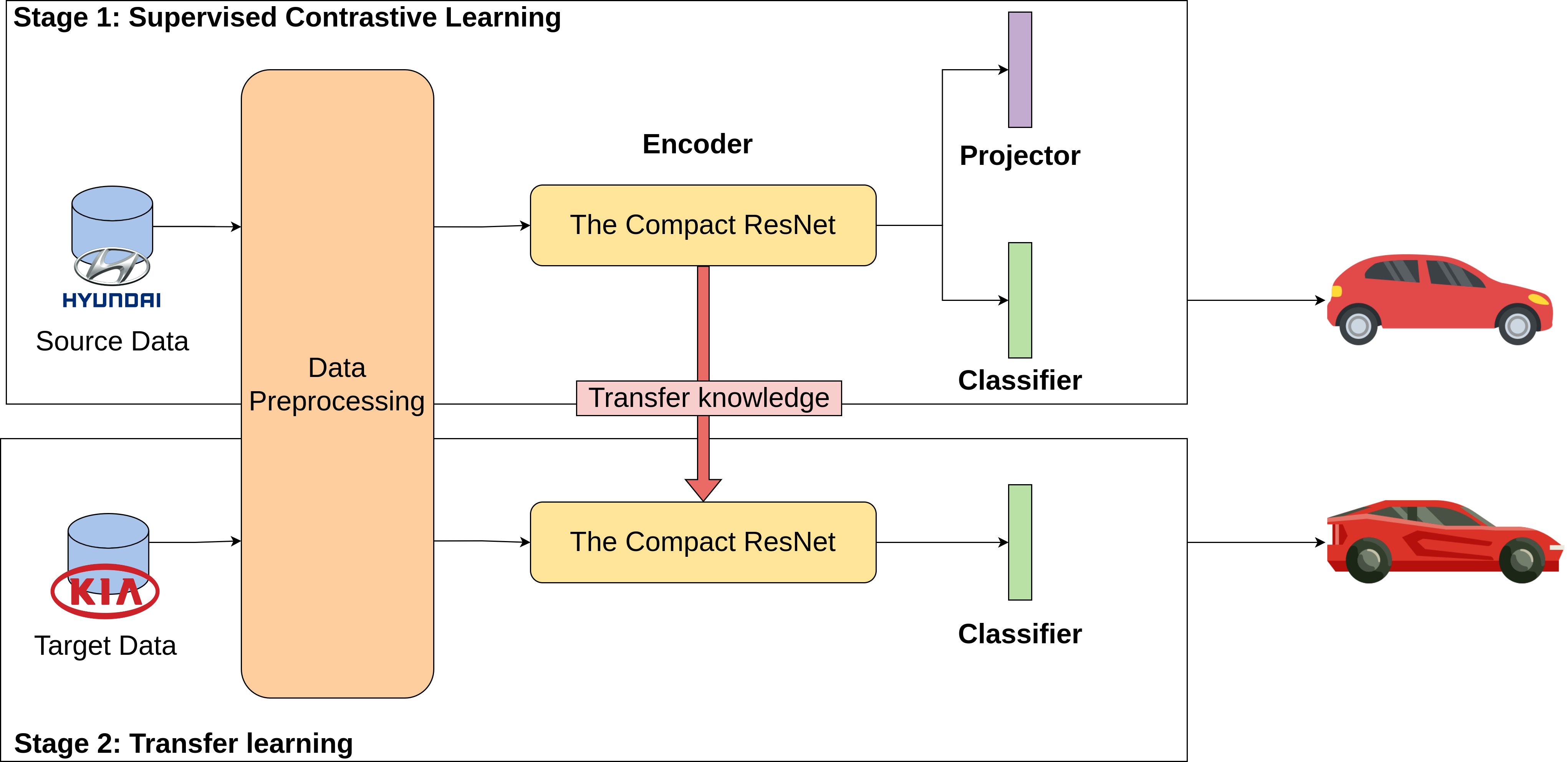}   
\caption{Overview of the proposed system.}
\label{fig:method_overview}
\end{figure*}

\subsection{Problem Formulation}

The intrusion detection system is installed in the CAN bus to monitor the CAN network and provide an alert if there is any malicious message. By using the fluctuation in the CAN ID sequence of consecutive CAN messages, this study aims to build an intrusion detection system $f$ that classifies a sequence of CAN IDs into a set of $C$ attack types, denoted as $\mathcal{A} = \{a_0, a_1, a_2, .., a_C\}$, where $a_0$ is a normal sequence. Suppose that several CAN log messages are collected from a car model under both normal and attack circumstances. After processing, the data $D = \{(X_i, y_i)\}$ are ready to train a machine learning model for the intrusion detection task, where $X_i$ is a sequence of CAN IDs and $y_i \in \mathcal{A}$ is the corresponding label. The objective is to build a multiple class classification IDS $f(\cdot|X)$, which determines whether a CAN IDs sequence $X_i$ belongs to a normal class or one of $C$ predefined attack types as accurately as possible.

The most important challenge for a supervised-based IDS is how to collect enough attack samples, especially for a newly released car model. We cannot apply a model trained on a CAN dataset A to another CAN dataset B, if A and B are from different manufacturers because each car model owns a specific CAN IDs pattern. Transfer learning can address this problem by extracting meaningful knowledge from a pre-trained model. We assume that an elegant source dataset $D_s$ exists collected from a car model $s$. From this, we can build an efficient IDS $f_s(\cdot|X_s)$ for the source car. The final objective is to build an IDS model for the target model $f_t(\cdot|X_t, f_s)$, with a limited target dataset $D_t$ where $|D_t| \ll |D_s|$.


\subsection{Proposed System}

The proposed system presented in \cref{fig:method_overview} comprises two stages: supervised contrastive learning from the source data and transfer learning for the target data. While supervised contrastive learning reduces the effect of class imbalance in the training data, transfer learning addresses the problem of limited data on a newly released car model.

In the first stage, the source data are fed into the SupCon ResNet model after being preprocessed. The proposed SupCon ResNet model includes three parts: the encoder, projection, and classification networks.
   \begin{itemize}
        \item The encoder network (Encoder) maps the input $X$ into the representation feature space. In this study, we designed a compact ResNet-18 architecture used as the encoder network. 
        \item The projection network (Projector) projects the outputs from the encoder network into another space, which is normalized and used to measure the distance for the supervised contrastive loss. From the experimental results, the model with a projector produces a higher accuracy, compared to the model without it \cite{Chen2020ARepresentations}. In addition, a non-linear projection network is assumed to be a better choice than the linear one. Hence, a multiple layer perception (MLP) with one hidden layer is used in our model.    
        \item The classification network (Classifier) is a linear layer performing the final classification task that maps the outputs from the encoder network to $C$ classes. 
    \end{itemize}
The classifier is trained with traditional CE-entropy loss thereafter, while the encoder and the project networks are trained together with SupCon loss.

In the second stage, we performed data processing on the target data the same as the source data. Training the target model from scratch may lead to the overfitting problem because of the small size dataset. Hence, we utilized the pre-trained model $f(s)$ for fine-tuning the target model. After completing the training of the source model, we transferred the weights of the encoder from the source model to the target model. After training the classifier on top of the frozen encoder, the entire model is trained with a small learning rate and the unfrozen encoder. The outcome of the system is the IDS model composed of the encoder and classification networks, which will be deployed in a real environment.


\subsection{Data Preprocessing}

In this study, only CAN IDs in the CAN messages were used, which was motivated by the message transmission mechanism in the CAN bus protocol \cite{Song2020In-vehicleNetwork}. Each CAN ID in the CAN bus has its message transmission cycle. Meanwhile, the message is broadcasted on the CAN bus based on the priority mechanism. Therefore, we believe that a certain sequential pattern exists in the CAN ID sequence. If an attacker injects some messages into the CAN bus, the pattern will be broken. We stacked $N$ sequential CAN IDs representing 29-bits, to create a CAN ID sequence frame. We choose $N = 29$ because of the convenience of processing square matrix in the CNN architecture network. Consequently, the input fed into the proposed model is a matrix of $29 \times 29$, which is visualized in \cref{fig:method_supcon}.

\subsection{ResNet Architecture}

\begin{figure}[t]
\centering
\includegraphics[width=2.5in]{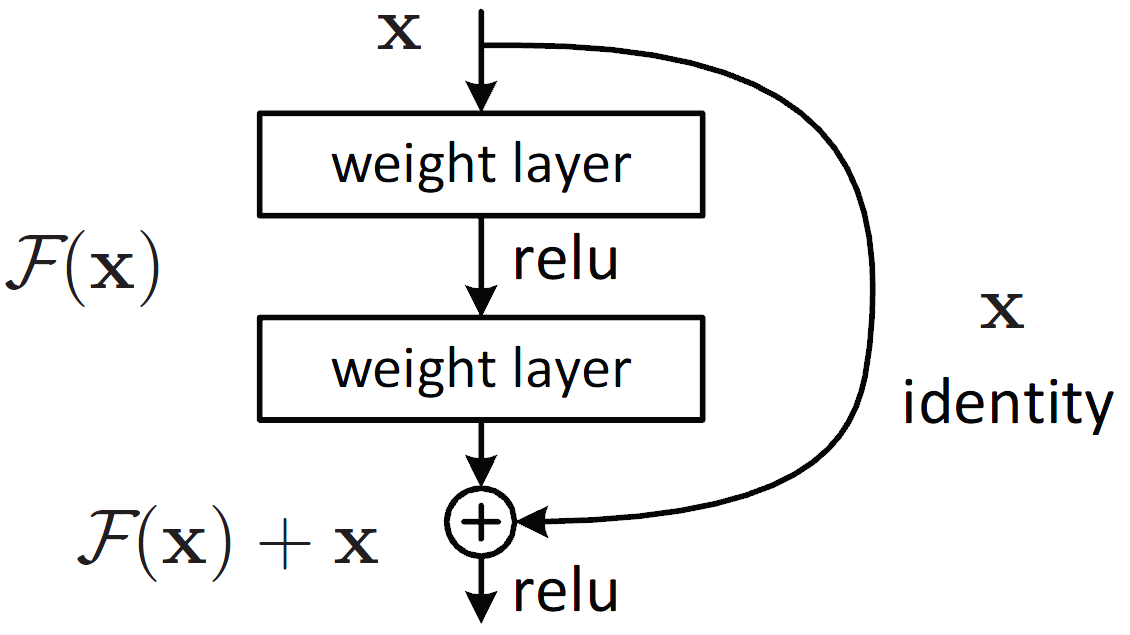}   
\caption{The residual block in ResNet}
\label{fig:method_skipconnection}
\end{figure}

The ResNet (Residual Network) \cite{He2015DeepRecognition}, which is introduced first in 2015 by He \textit{et al.}, has become a popular architecture in deep learning. This allows us to train a much deeper network without performance degradation. Theoretically, the more layer, the better result a deep learning model provides. However, in reality, training a deep network is challenging because of vanishing gradient and saturated training error. In a previous study regarding ResNet, the authors introduced the concept of skip connection to solve the problems. At that time, the 152-layer ResNet architecture outperformed the other well-known models, such as AlexNet and VGG, to win first place in the 2015-ILSVRC competition. 

The \cref{fig:method_skipconnection} displays a residual block, the core element in the ResNet. The block takes the input $\mathbf{x}$ and outputs $\mathbf{y}$ as follows:
\begin{equation}
    \mathbf{y} = \mathcal{F}(\mathbf{x}, \{W_i\}) + \mathbf{x}
\end{equation}
where the function $\mathcal{F}(\mathbf{x}, \{W_i\})$ represents the residual mapping to be learned with the weight of layers $W_i$. The $\mathbf{x}$ identity mapping is called a ``shortcut connection" that is element-wise addition to the function $\mathcal{F}$. Consequently, the simple concept of adding the identity $\mathbf{x}$ to the outcome does not increase the total number of parameters, depth, width, and computational cost (except for the negligible element-wise addition), compared to the plain network. The authors also showed that ResNet, which was 20 and 8 times deeper than AlexNet and VGG respectively, still has lower complexity. \cite{}

\begin{figure}[t]
\centering
\includegraphics[width=2in]{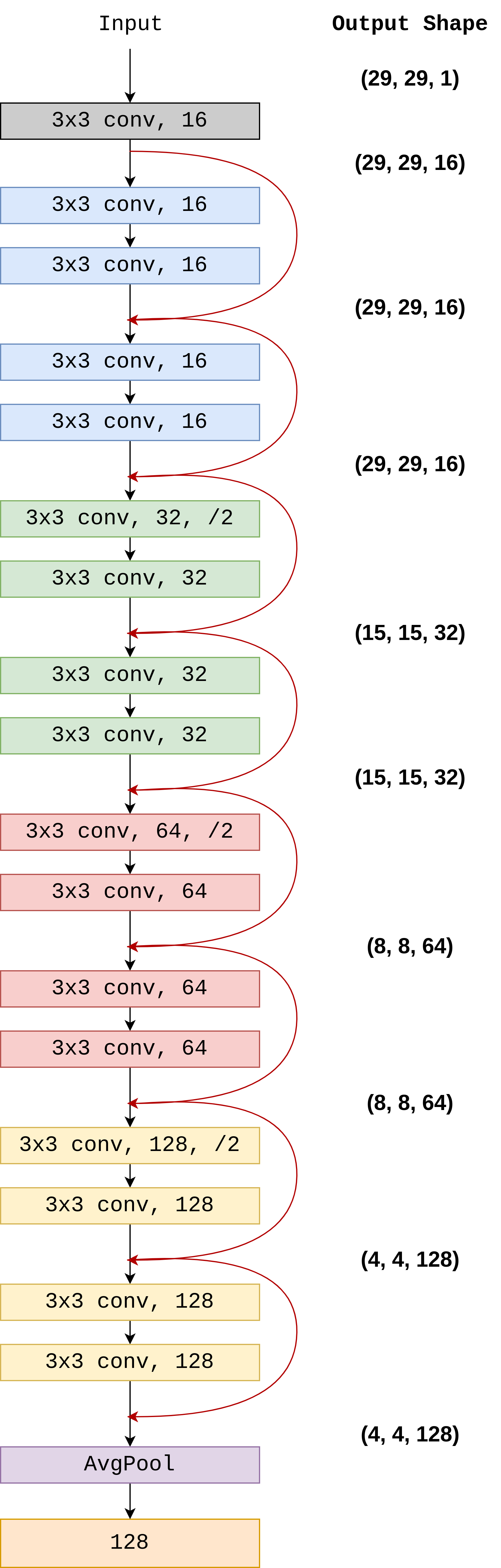}   
\caption{Detail of the compact ResNet encoder network}
\label{fig:method_resnet_arch}
\end{figure}

The CAN bus IDS requires not only high accuracy but also low running time. Therefore, we design a compact ResNet-18 to meet these requirements. The \cref{fig:method_resnet_arch} shows the detail of our architecture used for the encoder network. Specifically, we reduced the number of channels of each layer in the original ResNet-18 architecture by four times to produce the compact version. After passing through the convolutional layer with a kernel size of three, the sample goes through eight residual blocks to produce a hidden features representation with the size of 128.

\subsection{Supervised Contrastive Model}

\begin{figure}[t]
\centering
\includegraphics[width=3in]{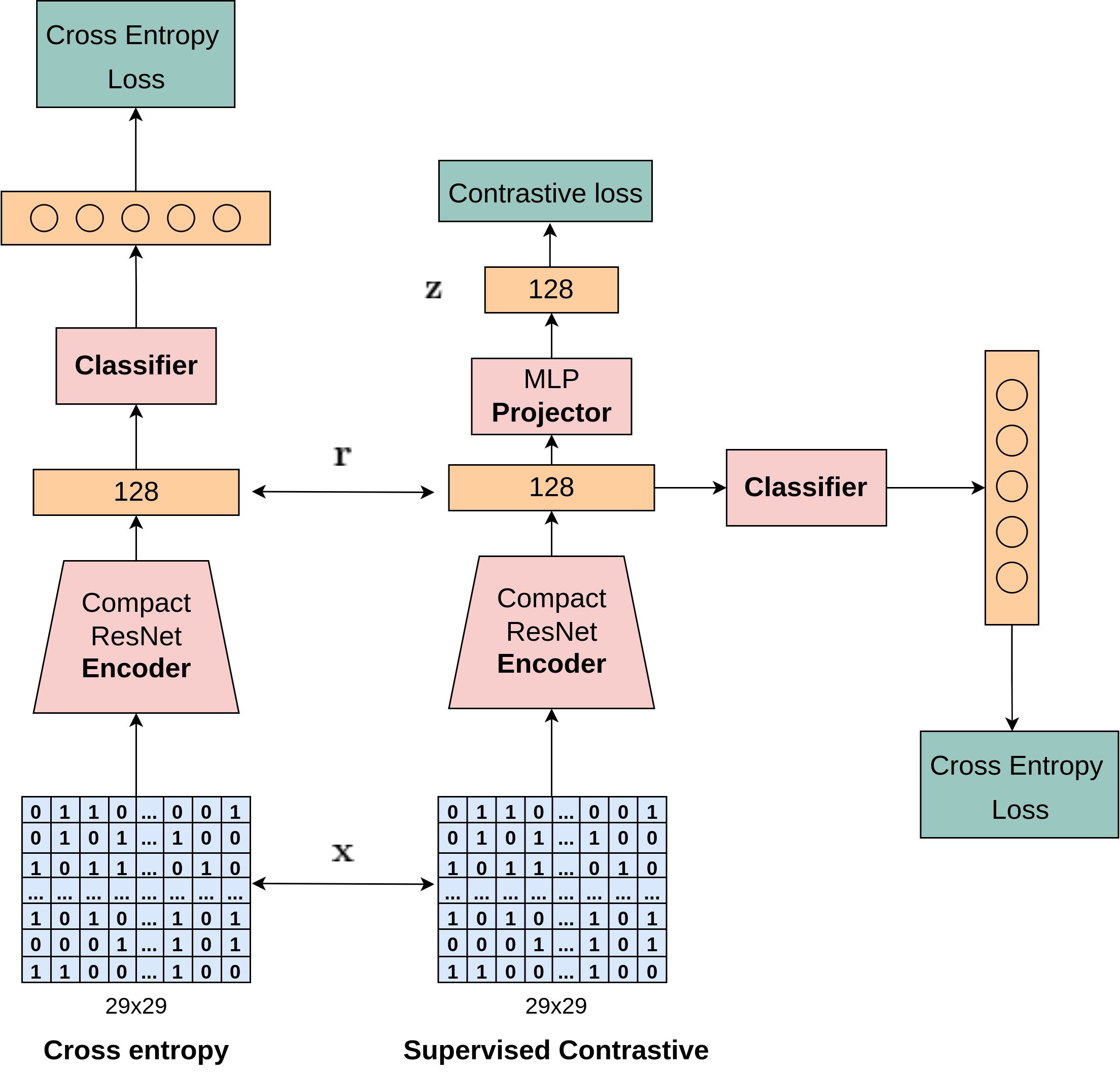}   
\caption{Difference between the SupCon ResNet and traditional CE ResNet models}
\label{fig:method_supcon}
\end{figure}

Compared to the traditional cross entropy that learns from the label, contrastive learning learns from the dissimilarity between samples. In general, the goal of contrastive learning is to produce embedding features wherein similar samples stay close to each other, while dissimilar ones are distant from each other. The original contrastive loss attempts to minimize the embedding distance when they are from the same class by taking a pair of $(\mathbf{x}_i, \mathbf{x}_j)$ as the input; otherwise, it maximizes the distance:
\begin{multline}
    \mathcal{L}_{\text{cont}}(\mathbf{x}_i, \mathbf{x}_j, \theta) = \mathbf{1}[y_i = y_j]||f_\theta(\mathbf{x}_i) - f_\theta(\mathbf{x}_j)||^2_2 \\ + \mathbf{1}[y_i \neq y_j]\text{max}(0, \epsilon -||f_\theta(\mathbf{x}_i) - f_\theta(\mathbf{x}_j)||^2_2)
\end{multline}
where $y_i$, $y_j$ are the labels of $\mathbf{x}_i$ and $\mathbf{x}_j$ respectively; $f_\theta(.)$ is the encoder function; and $\epsilon$ is a hyperparameter defining the lower bound distance between samples of different classes. Contrastive loss can be better than the cross-entropy if it is trained with hard positive/negative pairs. For example, a hard positive pair are two samples that belong to the same class but appear different. Meanwhile, a negative pair are those from different classes but appear quite similar. The challenge of how to design an efficient hard mining technique for training contrastive loss has still been being developed. In 2020, the SupCon loss \cite{Khosla2020SupervisedLearning} was proposed to solve the problem of the original contrastive loss. It outperformed the traditional cross-entropy on the ImageNet datasets in terms of the accuracy. The SupCon loss is extended from the contrastive learning in a self-supervised manner \cite{Chen2020ARepresentations} by leveraging the label information. \cref{fig:method_supcon} illustrates the structure of the supervised contrastive model, compared to conventional neural networks trained with cross-entropy loss. The SupCon architecture has three main parts: the encoder network, projector network, and classifier. The encoder network maps the input $\mathbf{x}$ to a representation vector, $\mathbf{r} = Enc(\mathbf{x})$. After being normalized, $\mathbf{r}$ is fed into the projector network and produces the $\mathbf{z} = Proj(\mathbf{r})$. The encoder and projector network are trained with the supervised contrastive loss as below:
\begin{multline}
    \mathcal{L}_{\text{supcon}} = -\sum_{i=1}^N \frac{1}{N_{y_i} - 1} \times \\\Bigg[\sum_{j = 1}^N \mathbf{1}_{i \neq j} \mathbf{1}_{y_i=y_j} \text{log} \frac{\text{exp}(\mathbf{z}_i \cdot \mathbf{z}_j/\tau)}{\sum_{k = 1}^N \mathbf{1}_{k \neq i}\text{exp}(\mathbf{z}_i \cdot \mathbf{z}_k/\tau)} \Bigg]
\end{multline}
where $N$ is the total number of training samples, $N_{y_i}$ is the number of positive samples that have the same label $y_i$ with the sample $i$, $j$ and $k$ are the indices of positive samples and all samples in the training set, respectively. In addition, temperature parameter $\tau$ controls the smoothness of probability distribution. A small $\tau$ will be good for training, but too small $\tau$ can lead to unstable training because of numerical instability. In the original paper, the authors state that $\mathcal{L}_{supcon}$ possesses the implicit property that encourages the hard positive/negative mining without performing it explicitly. As a result, the learning process for the supervised contrastive model does not need to perform hard mining, which slows down the training time. For more detail, we refer to the proof of the original paper \cite{Khosla2020SupervisedLearning}. By learning from dissimilarity/similarity, the model produces good representations $\mathbf{r}$, which are then used to train the classifier with cross-entropy loss as usual. 


\subsection{Transfer Learning}

Transfer learning is a technique for improving the performance of a target model on a limited target data by utilizing knowledge from a related source model, trained on an abundant source data. According to \cite{Pan2010ALearning}, the formal definition of transfer learning is described as follows:
\begin{itemize}
    \item Domain: A domain $\mathcal{D} = \{\mathcal{X}, P(X)\}$ includes two components: a feature space $\mathcal{X}$ and a marginal distribution $P(X)$ of data $X$. 
    \item Task: A task $\mathcal{T} = \{\mathcal{Y}, f\}$ comprises a label space $\mathcal{Y}$ and a predictive function $f = P(Y|X)$, which is trained on a dataset consists of pairs $\{x_i, y_i\}$, where $x_i \in X$ and $y_i \in Y$. 
    \item Transfer learning: Given a source domain $\mathcal{D}_s$, a corresponding source task $\mathcal{T}_s$, and a target domain $\mathcal{D}_T$ and target task $\mathcal{T}_T$, the objective of transfer learning is to learn the target conditional probability distribution $P(Y_T|X_T)$ in $\mathcal{D}_T$ with the information gained from $\mathcal{D}_S$ and $\mathcal{T}_S$. Here, $\mathcal{D}_S \neq \mathcal{D}_T$ or $\mathcal{T}_S \neq \mathcal{T}_T$. In most cases, a sufficient number of labeled source examples, which is extremely larger than the number of labeled target examples, are assumed to be available.
\end{itemize}
Based on the relationship between the source and target domains, source and target tasks, and availability of data, transfer learning techniques have various types: inductive,  transductive, and unsupervised. For more details on these categories, we refer to the survey of transfer learning \cite{Pan2010ALearning}.

We have the $\mathcal{D}_S$ and $\mathcal{D}_T$ domains from different car models to apply the transfer learning definition to the IDS for the CAN bus. Although they have the same data format that follows the predefined CAN messages structure, each dataset owns a different message transmission behavior. This implies the $\mathcal{X}_S = \mathcal{X}_T$, whereas $P(X_S) \neq P(X_T)$. These properties prevent the ability of a generalized IDS model for all car models. In practice, it is difficult to collect a large amount of training data for each car model. Based on the framework of \cite{Pan2010ALearning}, the domain-adaptation inductive transfer learning technique is suitable for our problem. We utilized the encoder in the supervised contrastive learning model, which is trained on a source data, as a pre-trained model. The predictive target model includes the encoder that has the same structure as the source model and top classifier, which is attached to solve the target task. First, we trained the target model with the frozen encoder. This step can be considered as the initialization of the weights of the classifiers. Finally, we trained the entire model after unfreezing the encoder network with a small learning rate to avoid overfitting.
\section{Experimental Results \label{sec:results}}

\subsection{Experiment Setup}

\begin{table}[t]
\caption{Number of messages in car hacking and survival datasets.}
\label{tab:datasize}
\begin{threeparttable}
    \begin{tabular}{c|L|L|L}
         \hline \textbf{Attack type} & \textbf{\#Messages (Sonata)} & \textbf{\#Messages (Soul)} & \textbf{ \#Messages (Spark)} \\
         \hline DoS Attack & 3,078,250 \break \textcolor{red}{587,521} & 181,901 \break \textcolor{red}{33,141} & 120,570 \break \textcolor{red}{22,587}\\ 
         \hline Fuzzy Attack & 3,347,013 \break \textcolor{red}{491,847} & 249,990 \break \textcolor{red}{39,812} & {65,665 \break \textcolor{red}{5,812}} \\ 
         \hline Malfunctioning & - & 173,436 \break \textcolor{red}{7,401} & 79,787 \break \textcolor{red}{8,047} \\ 
         \hline Gear Spoofing & 4,443,142 \break \textcolor{red}{597,252} & - & - \\ 
         \hline RPM Spoofing & 4,621,702 \break \textcolor{red}{654,897}& - & -\\
         \hline
    \end{tabular}
     \begin{tablenotes}
      \small
      \item Note: The Sonata values come from the car hacking dataset, while the other values are from the survival dataset. The red color indicates the injected messages, while the black indicates the normal messages. 
    \end{tablenotes}
\end{threeparttable}
\end{table}
%

We used two popular datasets produced by the Hacking and Countermeasure Research Lab (HCRL) of Korea University: the car hacking \cite{Song2020In-vehicleNetwork} and survival \cite{Han2018AnomalyAnalysis} datasets. The car hacking dataset was collected from the Hyundai Sonata model, and the survival dataset was collected from three specific car models (i.e., Hyundai Sonata, KIA Soul, and Chevrolet Spark). We used the data from the Sonata Hyundai car model in the car hacking dataset as the source and treated the others as the target ones since the data of the Hyundai Sonata car model is much larger than the others. The label space of source and target tasks are different: $\mathcal{Y}_S \in \mathbb{R}^5$ including normal, DoS, fuzzy, spoofing RPM, and spoofing gear information. Meanwhile, $\mathcal{Y}_T \in \mathbb{R}^4$, including normal, DoS, fuzzy, and malfunction. \cref{tab:datasize} lists the details of the data size.

Each message in both two datasets follows the same format containing a timestamp, CAN ID in HEX, the number of data bytes, 8 bytes of data, and the label. We extracted CAN IDs and transformed them from hexadecimal to 29-bit representation. Then, the CAN ID sequence frame was built by stacking 29 sequential samples together. A stride value $s$ is the number of messages between two continuous frames. More frames are created if a stride value is a small but less variant between these frames. We choose $s=15$ for the car hacking dataset and $s=10$ for the survival dataset because the size of the survival dataset is quite smaller than that of the car hacking dataset.

The dataset comprises multiple files corresponding to the attack types. For each file, the frame was labeled as normal if there is no injected message, whereas the label of the frame is a non-zero integer number considering the type of attack. We combined all of the processed frames to obtain the final dataset. The train/test splitting rate is 7:3. We trained the model on the training set and evaluated it on the test set. For each model, we performed the entire process from dataset splitting to evaluation five times and reported the average results.

The experiments were conducted on a server provided with 32 Intel(R) Xeon(R) Silver 4108 CPUs @ 1.80 GHz, a memory of 128 GB, and an Nvidia Titan RTX 24GB GPU. All of the code written in Python 3.7.10 and Pytorch 1.9.0 is published on the Github\footnote{Source code is available at https://github.com/htn274/CAN-SupCon-IDS}.


\subsection{Evaluation Metrics}

As our problem is multiclass classification, we evaluated the proposed model using the false-negative rate (FNR), recall (Rec), precision (Prec), and F1 score (F1). These metrics were calculated from the confusion matrix using true/false positive/negative samples. When we calculated a metric of a specific class, this class is considered positive, whereas the others are considered negative. Concretely, the FNR is the fraction between the false-negative and the total number of positive.
\begin{equation}
    \text{FNR} = \frac{FN}{TP + FN}
\end{equation}
In the case of the IDS, the FNR must be as small as possible. In addition, the FNRs of the attack classes are expected to be smaller than of the normal class because the consequence of a missed detected attack sample is more dangerous than that of a false alarm alert. Meanwhile, the recall is calculated as the fraction between true positive and the total sum of true positive and false negative. 
\begin{equation}
    \text{Rec} =  \frac{TP}{TP + FN}
\end{equation}
The precision is calculated as the fraction between the true positive and the total sum of true and false positives. 
\begin{equation} 
    \text{Pre} = \frac{TP}{TP + FP}
\end{equation}
For our problem, the proposed model should achieve both high recall and high precision. Therefore, the F1 score is used to balance these two metrics. The F1 score is defined as the harmonic mean of precision and recall.
\begin{equation}
    \text{F1}  = 2\times \frac{Prec \times Rec}{Prec + Recall}.
\end{equation}

\subsection{Choosing Hyperparameters for the SupCon ResNet Model}

\begin{figure*}[t]
\centering
\captionsetup{justification=centering}
\includegraphics[width=5in]{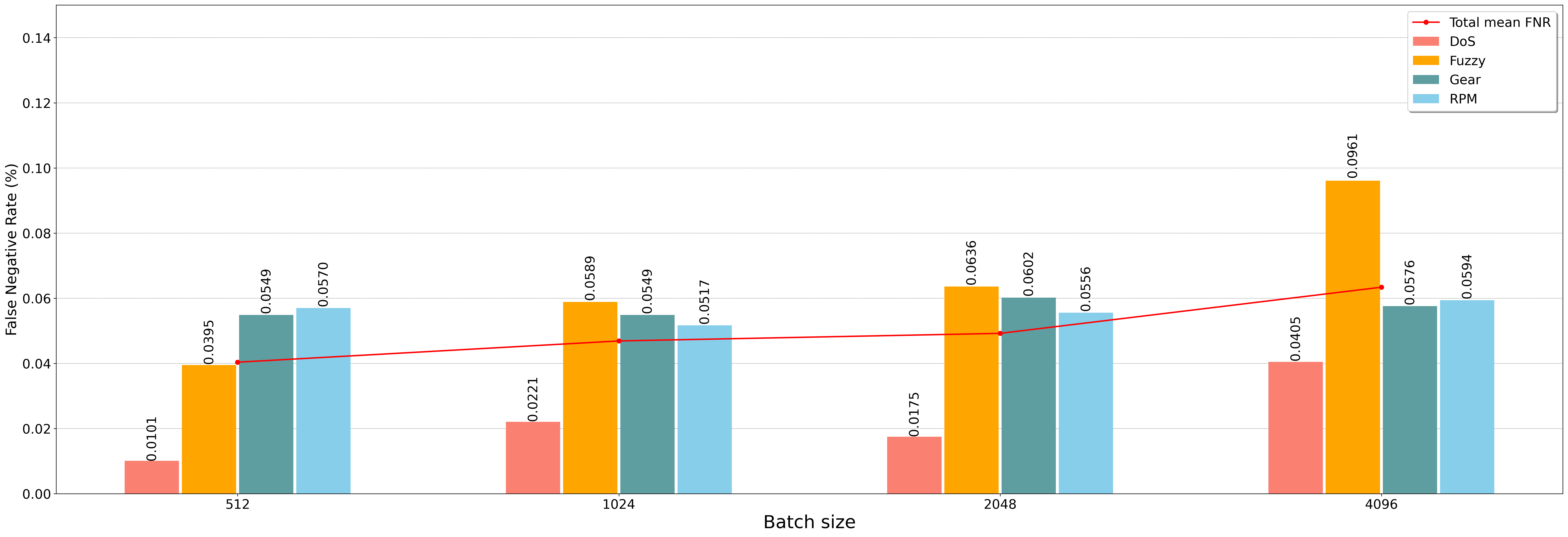}
\caption{False negative errors grouped by attack types of each set of hyperparameters}
\label{fig:results_hyperparams}
\end{figure*}

Choosing the right hyperparameters is important in deep learning because it can boost the performance of a model. Three hyperparameters should be considered in the SupCon ResNet model: the learning rate, batch size, and $\tau$ value. For simplicity, we set the $\tau$ value as 0.07 as advised in the original paper \cite{Khosla2020SupervisedLearning} and tuned only the learning rate and batch size. The authors of the SupCon study stated that the SupCon loss takes advantage of large batch sizes. Hence, we tried different batch sizes of 512, 1024, 2048, and 4096. We set the learning rate as 0.05 for 512 and 1024 batch sizes. Meanwhile, the learning rate of 0.1 was used for the model with the batch size of 2048 and 4096. Finally, we trained the top classifier for four models with the same configurations with 256 batch sizes and a learning rate of 0.01 after 150 epochs. We reported the false negative errors grouped by attack types of four models in \cref{fig:results_hyperparams}. The results showed that the classifier trained with SupCon loss classifies normal and attack samples well on the Car Hacking dataset, as all models have false-negative errors lower than 0.1\%. The Fuzzy attack has the highest false-negative errors in almost settings, while the DoS attack is easily detected. This finding is aligned with the study in \cite{Song2020In-vehicleNetwork}. Moreover, we compared these models in terms of the average false-negative errors. The results suggest that the larger batch size does not show better results. Overall, the model with a batch size of 512, which is the best model with the lowest mean of false-negative errors, is chosen for the following experiments. 

\subsection{Comparison with other Supervised Methods}

We compared the proposed SupCon ResNet model to two other models: the Inception ResNet \cite{Song2020In-vehicleNetwork}, which is the first study that demonstrates the concept of using CNN for CAN bus messages data, the baseline with the same compact ResNet encoder but using the cross-entropy loss. Regarding the Inception ResNet, we reimplemented the model with the hyperparameters reported in the study. Meanwhile, the ResNet trained with the cross-entropy loss was trained with a batch size of 256 and a learning rate of 0.001. \cref{tab:cmp_models} summarizes the results of these models, which are the average taken from five times running. 
 
 Although the CE ResNet obtains the highest false-negative error of the DoS attack, at 0.21\%, those of the other attacks are slightly lower than those of the Inception ResNet, by 0.06\%, 0.08\%, 0.09\% for the fuzzy attack, gear spoofing attack, and RPM spoofing attack, respectively. Surprisingly, the results of the SupCon ResNet outperform the others. Based on \cref{tab:cmp_models}, the SupCon ResNet reduces the false-negative errors of all attacks almost six times on average. In addition, the proposed model achieves the highest F1 score for all of the attacks, roughly 0.9998 on average.
  
\begin{figure*}[ht]
     \centering
     \subfloat[][CE ResNet model]{\includegraphics[width=3in]{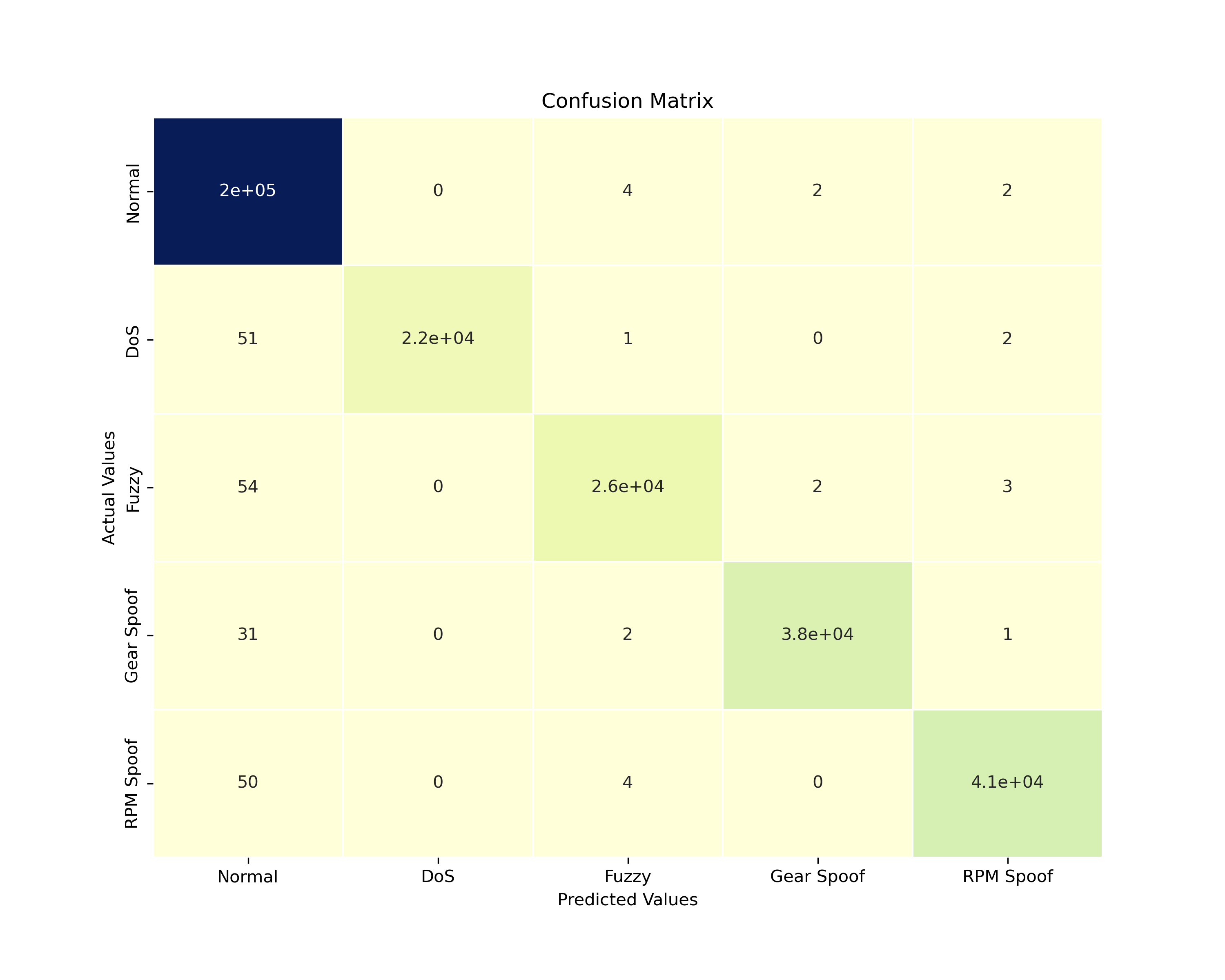}\label{<figure1>}}
     \subfloat[][SupCon ResNet model]{\includegraphics[width=3in]{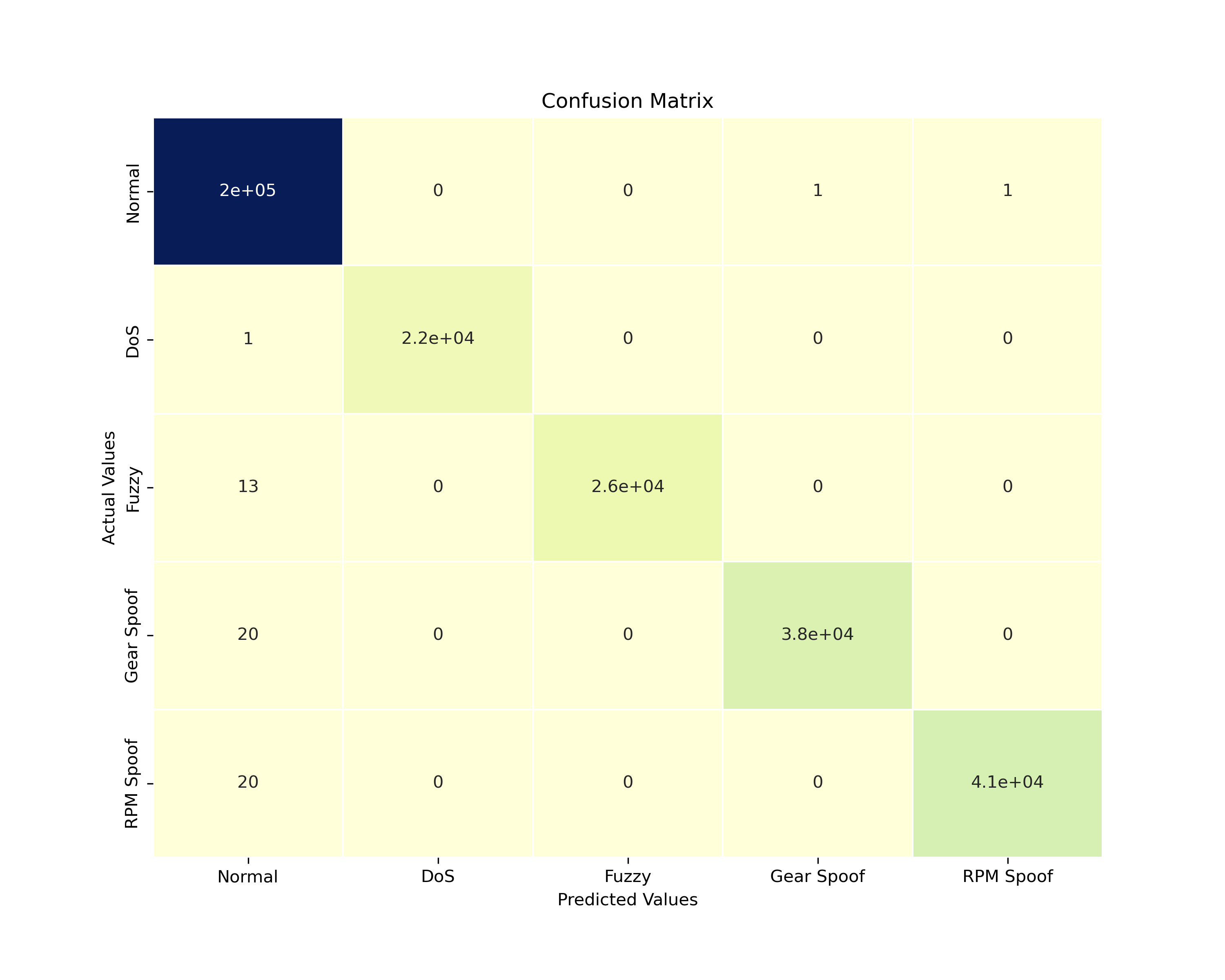}\label{<figure2>}}
     \caption{Comparison of confusion matrix between CE and SupCon models}
     \label{fig:confusion_matrix}
\end{figure*}

\cref{fig:confusion_matrix} illustrates the confusion matrices of the two models to show the superior of the SupCon ResNet compared to the baseline model. By learning from (dis)similarity, the SupCon ResNet decreases more than half of the number of miss-detected attack samples. There are 51 DoS samples classified falsely as a normal class in the case of the CE one, whereas there is only 1 sample in the case of the SupCon ResNet. Moreover, the SupCon ResNet also slightly decreases the false-negative samples within the attack classes and the number of false alarm cases.

In summary, these results indicate that the SupCon ResNet model separates the classes well on the car hacking dataset. This is because of the motivation of contrastive learning: the similar samples are pulled together while the dissimilar samples are pushed far apart. Due to the richness of the car hacking dataset, the SupCon ResNet can learn good representations, which are useful for classification, not only on the car hacking dataset but also on other CAN datasets, which will be proved empirically in the next section.

\begin{table}[ht]
    \centering
    \caption{Comparison of the proposed SupCon model \\and other supervised models}
    \begin{tabular}{L|c|c|c|c|c}
         \hline \textbf{Attack Type} & \textbf{Model} & \textbf{FNR} & \textbf{Rec} & \textbf{Prec} & \textbf{F1}  \\
         \hline 
         \multirow{3}{*}{DoS} & Inception ResNet &  0.12\% & 0.9988 & 0.9998& 0.9993 \\ 
          & CE ResNet & 0.21\% & 0.9979 & 0.9997 & 0.9987 \\ 
          & SupCon ResNet & \textbf{0.02\%} & \textbf{0.9998} & \textbf{1.0} & \textbf{0.9999} \\
         \hline 
         \multirow{3}{*}{Fuzzy} & Inception ResNet &  0.30\% & 0.9970 & 0.9982 & 0.9976 \\ 
          & CE ResNet &  0.24\% & 0.9976 & 0.9997 & 0.9987 \\ 
          & SupCon ResNet &  \textbf{0.06\%} & \textbf{0.9994} & \textbf{1.0} & \textbf{0.9997} \\
          \hline 
         \multirow{3}{*}{Gear Spoofing} & Inception ResNet &  0.19\% & 0.9981 & 0.9996 & 0.9988 \\ 
          & CE ResNet &  0.11\% & 0.9989 & 0.9997 & 0.9993 \\ 
          & SupCon ResNet &  \textbf{0.06\%} & \textbf{0.9994} & \textbf{0.9999} & \textbf{0.9997} \\
         \hline 
         \multirow{3}{*}{RPM Spoofing} & Inception ResNet & 0.21\% & 0.9979 & 0.9998 & 0.9988 \\ 
          & CE ResNet &  0.12\% & 0.9988 & 0.9998 & 0.9993 \\ 
          & SupCon ResNet &  \textbf{0.06\%} & \textbf{0.9994} & \textbf{1.0} & \textbf{0.9997} \\
         \hline 
    \end{tabular}
    \label{tab:cmp_models}
\end{table}


\subsection{Transfer Learning Results}

The SupCon ResNet produces good representations that contribute learned knowledge to train an efficient classification model using a small dataset. In this section, we will analyze the results of transfer learning from the proposed model, trained with the car hacking dataset, to other smaller-size and different car model datasets. We set up the experiment with three different configurations: 
\begin{enumerate}
    \item Random: We trained the model on the new dataset from scratch. This means that the weight of the model is randomly initialized.
    \item CE ResNet: We used the CE ResNet trained on the car hacking dataset as the pretrained model. 
    \item SupCon ResNet: We used the SupCon ResNet trained on the car hacking dataset as the pretrained model.
\end{enumerate}
The results of KIA Soul and Chevrolet Spark models are presented in \cref{tab:cmp_transfer_kia} and \cref{tab:cmp_transfer_spark}, respectively.

Overall, transfer learning reduces false-negative errors and increases the F1 score, especially for the fuzzy attack. For example, on both two target datasets, the false-negative errors of the fuzzy attack reduce by two times when using the CE ResNet as a pretrained model, compared to the random model. The number decreases significantly to 0.06\% and 0.47\% in the case of KIA Soul and Chevrolet Spark models, respectively, when the CE ResNet is replaced by the SupCon ResNet model. From the experiment results, we observe two important conclusions as follows:
\begin{itemize}
    \item For the IDS CAN domain, in the case of a small size dataset, transfer learning from a different car model can increase the detection abilities in terms of false-negative errors and F1 scores.
    \item Knowledge obtained from the SupCon ResNet model is better than that of the CE ResNet model. In fact, it increases the overall F1 score. 
\end{itemize}

\begin{table}[ht]
    \centering
    \caption{Transfer learning results for Kia Soul data}
    \begin{tabular}{c|c|c|c|c|c}
         \hline \textbf{Attack Type} & \textbf{Model} & \textbf{FNR} & \textbf{Rec} & \textbf{Prec} & \textbf{F1}  \\
         \hline 
        \multirow{3}{*}{Normal} & Random &  0.04\% & 0.9996 & 0.9980 & 0.9988 \\ 
          & CE ResNet & 0.04\% & 0.9996 & 0.9990 & 0.9993 \\ 
          & SupCon ResNet & \textbf{0.0\%} & \textbf{1.0} & \textbf{0.9999} & \textbf{0.9999} \\
         \hline 
         \multirow{3}{*}{DoS} & Random & 0.12\% & 0.9988 & 1.0 & 0.9994 \\ 
          & CE ResNet &  0.05\% & 0.9995 & 1.0 & 0.9997 \\ 
          & SupCon ResNet &  \textbf{0.01\%} & \textbf{0.9999} & \textbf{1.0} & \textbf{1.0} \\
         \hline 
         \multirow{3}{*}{Fuzzy} & Random & 0.49\% & 0.9951 & 0.9994 & 0.9973 \\ 
          & CE ResNet &  0.20\% & 0.9980 & 0.9992 & 0.9986  \\ 
          & SupCon ResNet & \textbf{0.06\%} & \textbf{0.9995} & \textbf{0.9999} & \textbf{0.9997} \\
          \hline 
         \multirow{3}{*}{Malfunction} & Random & 0.20\% & 0.9980 & 0.9973 & 0.9977\\ 
          & CE ResNet &  0.21\% & 0.9979 & 0.9984 & 0.9981 \\ 
          & SupCon ResNet & \textbf{0.06\%} & \textbf{0.9995} & \textbf{0.9999} & \textbf{0.9997} \\
         \hline 
         \multirow{3}{*}{Overall} & Random &  0.21\% & 0.9979 & 0.9987 & 0.9983 \\ 
          & CE ResNet &  0.13\% & 0.9988 & 0.9992 & 0.9989 \\ 
          & SupCon ResNet & \textbf{0.03\%} & \textbf{0.9997} & \textbf{1.0} & \textbf{0.9998}\\
         \hline 
    \end{tabular}
    \label{tab:cmp_transfer_kia}
\end{table}

\begin{table}[ht]
    \centering
    \caption{Transfer learning results for Chevrolet Spark data}
    \begin{tabular}{c|c|c|c|c|c}
         \hline \textbf{Attack Type} & \textbf{Model} & \textbf{FNR} & \textbf{Rec} & \textbf{Prec} & \textbf{F1}  \\
         \hline 
        \multirow{3}{*}{Normal} & Random & 0.25\% & 0.9975 & 0.9960 & 0.9967 \\ 
          & CE ResNet & 0.11\% & 0.9989 & 0.9982 & 0.9985 \\ 
          & SupCon ResNet & \textbf{0.11\%} & \textbf{0.9989} & \textbf{0.9992} & \textbf{0.9991}\\
         \hline 
         \multirow{3}{*}{DoS} & Random & 0.18\% & 0.9982 & 1.0& 0.9991 \\ 
          & CE ResNet &  0.05\% & 0.9995 & 1.0 & 0.9998 \\ 
          & SupCon ResNet & \textbf{0.04\%} & \textbf{0.9996} & \textbf{1.0}& \textbf{0.9998} \\
         \hline 
         \multirow{3}{*}{Fuzzy} & Random & 2.99\% & 0.9701& 0.9737& 0.9719 \\ 
          & CE ResNet &   1.46\% & 0.9854 & 0.9901 & 0.9878 \\ 
          & SupCon ResNet & \textbf{0.47\%}& \textbf{0.9953}& \textbf{0.9902}& \textbf{0.9927} \\
          \hline 
         \multirow{3}{*}{Malfunction} & Random & 0.28\% & 0.9972& 0.9994& 0.9983\\  & CE ResNet &  0.13\% & 0.9987 & 0.9989 & 0.9988 \\ 
          & SupCon ResNet & \textbf{0.04\%} & \textbf{0.9996}& \textbf{1.0}& \textbf{0.9998} \\
         \hline 
         \multirow{3}{*}{Overall} & Random & 0.93\% & 0.9908& 0.9923 & 0.9915 \\ 
          & CE ResNet & 0.44\% & 0.9956 & 0.9968 & 0.9962 \\ 
          & SupCon ResNet &  \textbf{0.17\%} & \textbf{0.9984}& \textbf{0.9974}& \textbf{0.9979} \\
         \hline 
    \end{tabular}
    \label{tab:cmp_transfer_spark}
\end{table}

\subsection{Model Complexity Analysis}

Along with high detection accuracy, a CAN-bus IDS must follow hardware constraints, such as low computing power, small memory size, and small response time, such that it can be installed and deployed in an ECU. \cref{tab:cmp_model_complexity} shows the complexity of the proposed model in terms of memory and running time, compared to the Inception ResNet proposed by \cite{Song2020In-vehicleNetwork}. From the results, our model is more lightweight and runs faster than the state of the art of CAN bus IDS. The model spends approximately 3.65 ms when inferencing one frame including 29 messages on a 64-bit Intel (R) Core(TM) i7-7700 CPU @ 3.6 GHz. This implies that the model can process up to approximately 8000 messages within a second. Simultaneously, there are approximately 2000 CAN messages on the bus. Therefore, our proposed model has enough capability to be deployed in a real ECU. 

\begin{table}[ht]
    \centering
    \caption{Comparison of model complexity}
    \begin{tabular}{c|L|l|l}
         \hline \textbf{Model} & \textbf{Total parameters (M)} & \textbf{Total size (MB)} &  \textbf{Inference  time} \break \textbf{(ms)} \\
         \hline Inception ResNet \cite{Song2020In-vehicleNetwork} & 1.69 & 9.33 & 4.49 \\ 
         \hline SupCon ResNet & 0.70 & 5.13 & 3.65 \\ 
         \hline 
    \end{tabular}
    \label{tab:cmp_model_complexity}
\end{table}


\section{Conclusion \label{sec:conclusion}}

This study aims to develop a lightweight and efficient CAN bus intrusion detection system capable of detecting and identifying specific types of attacks. We introduced a SupCon ResNet, which was trained with the supervised contrastive loss, by manipulating the CAN ID sequences in binary form. To the best of our knowledge, this is the first study that combined supervised contrastive learning and transfer learning to the in-vehicle IDS. The results of experiments on real datasets indicate that the supervised contrastive loss significantly reduced the false-negative rate, especially for the fuzzy attack. Consequently, a lightweight deep learning model, which occupies low memory size and runs faster, can be deployed in an ECU to serve the intrusion detection task without detection accuracy degradation. Moreover, the proposed model is useful for transfer learning, solving the challenge of data lacking in the new release car model. We showed that the transfer learning from the SupCon ResNet model is superior to other baselines. Concretely, the proposed model decreases the false-negative rate by 4 and 2.5 times when tested with KIA Soul and Chevrolet Spark models, respectively. From these results, we concluded that the SupCon ResNet can deal with the multiple attack classification on the CAN bus. Furthermore, the model saves time and labor costs for data collection for the new car model. However, the proposed system contains a limitation of no unknown attack detection. The model should be frequently updated to detect new types of attacks since the multiclass classification cannot be solved using unsupervised models. The challenge of adding a new attack type without training the model from the beginning can be a promising direction for future studies. In addition, the idea of personalized federating learning can be adopted to CAN IDS such that each vehicle can contribute its own data to build a competent global model for all vehicles.

\printbibliography

@article{DAngelo2020AVehicles,
    title = {{A cluster-based multidimensional approach for detecting attacks on connected vehicles}},
    year = {2020},
    journal = {IEEE Internet of Things Journal},
    author = {D'Angelo, Gianni and Castiglione, Arcangelo and Palmieri, Francesco},
    publisher = {Institute of Electrical and Electronics Engineers Inc.},
    doi = {10.1109/JIOT.2020.3032935}
}

@article{Liu2022ADetection,
    title = {{A multi-task based deep learning approach for intrusion detection}},
    year = {2022},
    journal = {Knowledge-Based Systems},
    author = {Liu, Qigang and Wang, Deming and Jia, Yuhang and Luo, Suyuan and Wang, Chongren},
    month = {2},
    pages = {107852},
    volume = {238},
    publisher = {Elsevier},
    doi = {10.1016/J.KNOSYS.2021.107852},
    issn = {0950-7051}
}

@article{Kang2016ASecurity,
    title = {{A novel intrusion detection method using deep neural network for in-vehicle network security}},
    year = {2016},
    journal = {IEEE Vehicular Technology Conference},
    author = {Kang, Min Ju and Kang, Je Won},
    month = {7},
    volume = {2016-July},
    publisher = {Institute of Electrical and Electronics Engineers Inc.},
    doi = {10.1109/VTCSPRING.2016.7504089}
}

@article{Chen2020ARepresentations,
    title = {{A Simple Framework for Contrastive Learning of Visual Representations}},
    year = {2020},
    journal = {37th International Conference on Machine Learning, ICML 2020},
    author = {Chen, Ting and Kornblith, Simon and Norouzi, Mohammad and Hinton, Geoffrey},
    month = {2},
    pages = {1575--1585},
    volume = {PartF168147-3},
    publisher = {International Machine Learning Society (IMLS)},
    url = {https://arxiv.org/abs/2002.05709v3},
    isbn = {9781713821120},
    doi = {10.48550/arxiv.2002.05709},
    arxivId = {2002.05709}
}

@article{Jo2021ACountermeasures,
    title = {{A Survey of Attacks on Controller Area Networks and Corresponding Countermeasures}},
    year = {2021},
    journal = {IEEE Transactions on Intelligent Transportation Systems},
    author = {Jo, Hyo Jin and Choi, Wonsuk},
    publisher = {Institute of Electrical and Electronics Engineers Inc.},
    doi = {10.1109/TITS.2021.3078740}
}

@article{Pan2010ALearning,
    title = {{A survey on transfer learning}},
    year = {2010},
    journal = {IEEE Transactions on Knowledge and Data Engineering},
    author = {Pan, Sinno Jialin and Yang, Qiang},
    number = {10},
    pages = {1345--1359},
    volume = {22},
    doi = {10.1109/TKDE.2009.191},
    issn = {10414347}
}

@article{Kang2021ASystem,
    title = {{A Transfer Learning based Abnormal CAN Bus Message Detection System}},
    year = {2021},
    journal = {Proceedings - 2021 IEEE 18th International Conference on Mobile Ad Hoc and Smart Systems, MASS 2021},
    author = {Kang, Liuwang and Shen, Haiying},
    pages = {545--553},
    publisher = {Institute of Electrical and Electronics Engineers Inc.},
    isbn = {9781665449359},
    doi = {10.1109/MASS52906.2021.00073}
}

@article{Avatefipour2019AnLearning,
    title = {{An intelligent secured framework for cyberattack detection in electric vehicles' can bus using machine learning}},
    year = {2019},
    journal = {IEEE Access},
    author = {Avatefipour, Omid and Saad Al-Sumaiti, Ameena and El-Sherbeeny, Ahmed M. and Mahrous Awwad, Emad and Elmeligy, Mohammed A. and Mohamed, Mohamed A. and Malik, Hafiz},
    pages = {127580--127592},
    volume = {7},
    publisher = {Institute of Electrical and Electronics Engineers Inc.},
    doi = {10.1109/ACCESS.2019.2937576}
}

@inproceedings{Taylor2016AnomalyNetworks,
    title = {{Anomaly detection in automobile control network data with long short-term memory networks}},
    year = {2016},
    booktitle = {Proceedings - 3rd IEEE International Conference on Data Science and Advanced Analytics, DSAA 2016},
    author = {Taylor, Adrian and Leblanc, Sylvain and Japkowicz, Nathalie},
    month = {12},
    pages = {130--139},
    publisher = {Institute of Electrical and Electronics Engineers Inc.},
    isbn = {9781509052066},
    doi = {10.1109/DSAA.2016.20}
}

@article{Han2018AnomalyAnalysis,
    title = {{Anomaly intrusion detection method for vehicular networks based on survival analysis}},
    year = {2018},
    journal = {Vehicular Communications},
    author = {Han, Mee Lan and Kwak, Byung Il and Kim, Huy Kang},
    month = {10},
    pages = {52--63},
    volume = {14},
    publisher = {Elsevier},
    doi = {10.1016/J.VEHCOM.2018.09.004},
    issn = {2214-2096}
}

@article{Andresini2021Autoencoder-basedDetection,
    title = {{Autoencoder-based deep metric learning for network intrusion detection}},
    year = {2021},
    journal = {Information Sciences},
    author = {Andresini, Giuseppina and Appice, Annalisa and Malerba, Donato},
    month = {8},
    pages = {706--727},
    volume = {569},
    publisher = {Elsevier},
    doi = {10.1016/J.INS.2021.05.016},
    issn = {0020-0255}
}

@book{1991BOSCH2.0,
    title = {{BOSCH CAN Specification Version 2.0}},
    year = {1991}
}

@article{He2015DeepRecognition,
    title = {{Deep Residual Learning for Image Recognition}},
    year = {2015},
    journal = {Proceedings of the IEEE Computer Society Conference on Computer Vision and Pattern Recognition},
    author = {He, Kaiming and Zhang, Xiangyu and Ren, Shaoqing and Sun, Jian},
    month = {12},
    pages = {770--778},
    volume = {2016-December},
    publisher = {IEEE Computer Society},
    url = {https://arxiv.org/abs/1512.03385v1},
    isbn = {9781467388504},
    doi = {10.48550/arxiv.1512.03385},
    issn = {10636919},
    arxivId = {1512.03385}
}

@inproceedings{Koscher2010ExperimentalAutomobile,
    title = {{Experimental Security Analysis of a Modern Automobile}},
    year = {2010},
    booktitle = {2010 IEEE Symposium on Security and Privacy},
    author = {Koscher, Karl and Czeskis, Alexei and Roesner, Franziska and Patel, Shwetak and Kohno, Tadayoshi and Checkoway, Stephen and Mccoy, Damon and Kantor, Brian and Anderson, Danny and Shacham, Hovav and Savage, Stefan},
    pages = {447--462},
    url = {http://www.autosec.org/}
}

@article{Cho2016FingerprintingDetection,
    title = {{Fingerprinting Electronic Control Units for Vehicle Intrusion Detection}},
    year = {2016},
    author = {Cho, Kyong-Tak and Shin, Kang G},
    pages = {911},
    url = {https://www.usenix.org/conference/usenixsecurity16/technical-sessions/presentation/cho},
    isbn = {978-1-931971-32-4}
}

@article{Seo2018GIDS:Network,
    title = {{GIDS: GAN based Intrusion Detection System for In-Vehicle Network}},
    year = {2018},
    journal = {2018 16th Annual Conference on Privacy, Security and Trust, PST 2018},
    author = {Seo, Eunbi and Song, Hyun Min and Kim, Huy Kang},
    month = {10},
    publisher = {Institute of Electrical and Electronics Engineers Inc.},
    doi = {10.1109/PST.2018.8514157}
}

@article{Derhab2021Histogram-BasedNetworks,
    title = {{Histogram-Based Intrusion Detection and Filtering Framework for Secure and Safe In-Vehicle Networks}},
    year = {2021},
    journal = {IEEE Transactions on Intelligent Transportation Systems},
    author = {Derhab, Abdelouahid and Belaoued, Mohamed and Mohiuddin, Irfan and Kurniawan, Fajri and Khan, Muhammad Khurram},
    pages = {1--14},
    url = {https://ieeexplore.ieee.org/document/9463874/},
    doi = {10.1109/TITS.2021.3088998}
}

@article{Song2020In-vehicleNetwork,
    title = {{In-vehicle network intrusion detection using deep convolutional neural network}},
    year = {2020},
    journal = {Vehicular Communications},
    author = {Song, Hyun Min and Woo, Jiyoung and Kim, Huy Kang},
    month = {1},
    pages = {100198},
    volume = {21},
    publisher = {Elsevier Inc.},
    doi = {10.1016/j.vehcom.2019.100198},
    issn = {22142096}
}

@article{Nam2021IntrusionNetworks,
    title = {{Intrusion Detection Method Using Bi-Directional GPT for in-Vehicle Controller Area Networks}},
    year = {2021},
    journal = {IEEE Access},
    author = {Nam, Minki and Park, Seungyoung and Kim, Duk Soo},
    pages = {124931--124944},
    volume = {9},
    publisher = {Institute of Electrical and Electronics Engineers Inc.},
    doi = {10.1109/ACCESS.2021.3110524},
    issn = {21693536}
}

@article{ChopraLearningVerification,
    title = {{Learning a Similarity Metric Discriminatively, with Application to Face Verification}},
    author = {Chopra, Sumit and Hadsell, Raia and Lecun, Yann}
}

@article{Ashraf2020NovelSystems,
    title = {{Novel Deep Learning-Enabled LSTM Autoencoder Architecture for Discovering Anomalous Events From Intelligent Transportation Systems}},
    year = {2020},
    journal = {IEEE Transactions on Intelligent Transportation Systems},
    author = {Ashraf, Javed and Bakhshi, Asim D. and Moustafa, Nour and Khurshid, Hasnat and Javed, Abdullah and Beheshti, Amin},
    month = {9},
    pages = {1--12},
    publisher = {Institute of Electrical and Electronics Engineers (IEEE)},
    doi = {10.1109/tits.2020.3017882},
    issn = {1524-9050}
}

@article{Desta2022Rec-CNN:Plots,
    title = {{Rec-CNN: In-vehicle networks intrusion detection using convolutional neural networks trained on recurrence plots}},
    year = {2022},
    journal = {Vehicular Communications},
    author = {Desta, Araya Kibrom and Ohira, Shuji and Arai, Ismail and Fujikawa, Kazutoshi},
    month = {6},
    pages = {100470},
    volume = {35},
    publisher = {Elsevier BV},
    url = {www.elsevier.com/locate/vehcom},
    doi = {10.1016/J.VEHCOM.2022.100470},
    issn = {22142096}
}

@article{Hoppe2011SecurityCountermeasures,
    title = {{Security threats to automotive CAN networks—Practical examples and selected short-term countermeasures}},
    year = {2011},
    journal = {Reliability Engineering {\&} System Safety},
    author = {Hoppe, Tobias and Kiltz, Stefan and Dittmann, Jana},
    number = {1},
    month = {1},
    pages = {11--25},
    volume = {96},
    publisher = {Elsevier},
    doi = {10.1016/J.RESS.2010.06.026},
    issn = {0951-8320}
}

@article{Khosla2020SupervisedLearning,
    title = {{Supervised Contrastive Learning}},
    year = {2020},
    author = {Khosla, Prannay and Teterwak, Piotr and Wang, Chen and Sarna, Aaron and Research, Google and Tian, Yonglong and Isola, Phillip and Maschinot, Aaron and Liu, Ce and Krishnan, Dilip},
    month = {4},
    url = {https://arxiv.org/abs/2004.11362v5},
    doi = {10.48550/arxiv.2004.11362},
    arxivId = {2004.11362}
}

@article{Lopez-Martin2022SupervisedDetection,
    title = {{Supervised contrastive learning over prototype-label embeddings for network intrusion detection}},
    year = {2022},
    journal = {Information Fusion},
    author = {Lopez-Martin, Manuel and Sanchez-Esguevillas, Antonio and Arribas, Juan Ignacio and Carro, Belen},
    month = {3},
    pages = {200--228},
    volume = {79},
    publisher = {Elsevier},
    doi = {10.1016/J.INFFUS.2021.09.014},
    issn = {1566-2535}
}

%

\end{document}